\documentclass[
]{ceurart}

\usepackage{listings}
\usepackage{caption}
\begin{document}

\copyrightyear{2024}
\copyrightclause{Copyright for this paper by its authors.
  Use permitted under Creative Commons License Attribution 4.0
  International (CC BY 4.0).}

\conference{ }

\title{Coca4ai: checking energy behaviors on AI data centers}

\tnotemark[1]

\author[1]{Paul Gay}[]
\author[2]{Éric Bilinski}[]
\author[2]{Anne-Laure Ligozat}[]
\address[1]{Université de Pau et des Pays de l'Adour}
\address[2]{Laboratoire LISN, université Paris-Saclay}

\begin{abstract}
Monitoring energy behaviors in AI data centers is crucial, both to reduce their energy consumption and to raise awareness among their users which are key actors in the AI field. This paper shows a proof of concept of easy and lightweight monitoring of energy behaviors at the scale of a whole data center, a user or a job submission. Our system uses software wattmeters and we validate our setup with per node accurate external wattmeters. Results show that there is an interesting potential from the efficiency point of view, providing arguments to create user engagement thanks to energy monitoring.
\end{abstract}


\maketitle

\section{Introduction}

The environmental footprint of artificial intelligence is a growing concern due to its recent democratization~\cite{couillet2022submerged,cowls2023ai,kaack2022aligning}. 
Mitigating strategies can rely on related work about environmental impact of ICT, with open source software powermeters ~\cite{jay2023experimental} to measure energy consumption at the scope of a program run, or more global methodologies which include Life Cycle Analysis (LCA)~\cite{itten2020digital}, indirect effects~\cite{rasoldier2022realistic} and other criteria such as water consumption~\cite{siddik2021environmental,li2023making}.
As studies for AI are emerging~\cite{berthelot2023estimating,lefevre2023environmental}, these findings needs 
to be integrated in industrial and research data centers practices. In particular, building from previous studies~\cite{khan2019analyzing}, we advocate that not only the environnment but practicioners can benefit from such studies to improve their efficiency at work. In other words, our hypothesis is that profiling user energy behavior is an interesting approach to draw the attention of users to their environmental footprint. Powermeter equipped data centers such as Grid'5000 platform~\cite{de2012green} can enable such behaviors. However, their use is based on a voluntary basis, thus reducing its scope to a subset of the users. 
Cloud solutions report carbon metrics, but might be bound to a virtual machine, thus losing the details regarding each task or job. 
Last but not least, aggregation of job statistics at a higher level is crucial to identify relevant energy behaviors. To the best of our knowledge, research interest on the topic of energy behavior has been mostly restricted to simulations~\cite{madon2022characterization}.
The current work aims at filling this gap by providing such a setup for research purposes. More precisely, we present a system to monitor energy behaviors deployed on the labia~\footnote{https://lab-ia.fr} data center. For each job, we record the use of GPUs and CPUs in terms of memory and use of calculation capacities, as well as the electrical power consumed as measured by Nvidia-smi and RAPL. These measurements are validated with accurate external wattmeters. The collected data enables us to identify new opportunities for data centers, namely under use of the GPUs and badly configured jobs.




\section{Methodology and presentation of the system}

\begin{figure}
\centering
 \includegraphics[width=.8\textwidth]{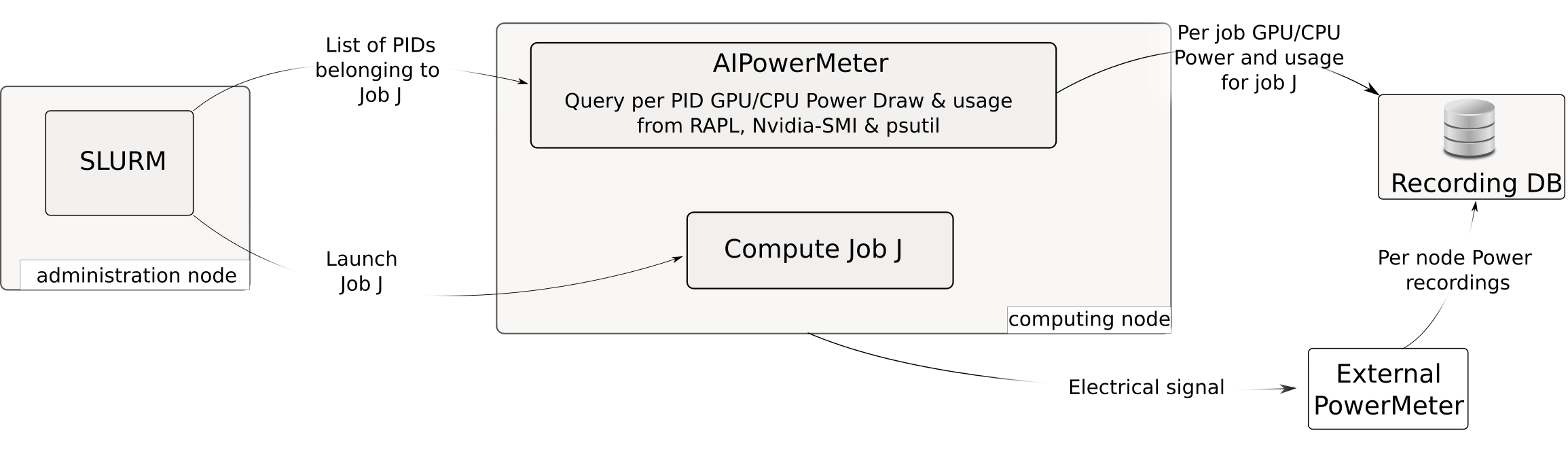} 
\caption{\textit{Global view of our setup to profile CPU and GPU usages with software and external powermeters. Usage and power draws are attributed to each PID corresponding to each job launched by SLURM.}}
 \label{fig}
\end{figure}

The SLURM based lab-ia cluster is composed of 12 nodes hosting 32 GPU Tesla K80 and Tesla V100. It is a small scale center designed to be used to develop prototypes and small scale experiments by researchers. We use Omegawatt~\footnote{\url{mv.omegawatt.fr/}} products as external wattmeter, to replace the power cables of each machine with extensions which include an intensity sensor connected to our recording database. 
RAPL and Nvidia data are recorded thanks to AIPowerMeter~\footnote{\url{https://greenai-uppa.github.io/AIPowerMeter/}} but other tools cited in the previous section could have been used. Data collection is summarized in figure~\ref{fig}. For each job $J$ launched by SLURM, a prolog program starts the AIPowerMeter software to collect GPU and CPU power consumption and usage. As multiple jobs can run on the same node, we regularly use the \textit{scontrol} SLURM tool to update the list of the Process Identifiers (PIDs) belonging to the job $J$ which allows us to assign the amount of power based on the relative cpu time used by its processes, as measured by \textit{psutil}. 
We compared the energy consumption from the software and external powermeters and found that on average, multiplying the software recording with a constant factor enables us to estimate the external wattmeter value within an error of 16\%. This gap can be attributed to the behaviors of the devices not monitored by RAPL (Hard disk, mother board, network interface).

\section{Observed energy behaviors}
\begin{figure}
\centering
 \includegraphics[width=.8\textwidth]{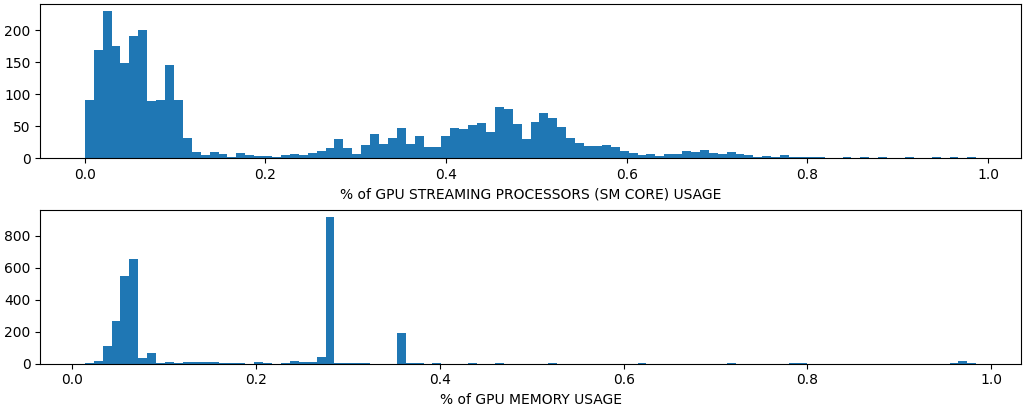} 
 \caption{\textit{Histograms of GPU SM cores and memory GPU usage. Two peaks are present in the distribution but none of the jobs is using the GPUs at their full capacity}}
 \label{fig:gpu}
\end{figure}

The global job statistics and GPU usage shown in this section are collected over 20 days in November 2023. Table~\ref{tab:slurm} shows energy consumption over the different job status. It can be noticed that there is a significant contribution of jobs which are FAILED (13\%), CANCELLED (5\%) and specially TIMEOUT (41\%), where the job was automatically stopped by SLURM. In the last case, this corresponds to a few very long jobs which ultimately represent a large portion of the total consumption. Overall, only 40\% of the power consumption corresponds to completed jobs, which is in line with previous studies showing the lack of efficiency of user behaviors~\cite{khan2019analyzing}.
 \begin{table}
\centering 
\begin{tabular}{| c | c | c | c | c|}
Status & \#JOBS & GPU (kWh) & CPU (kWh) & Ext. (kWh)\\
COMPLETED&  1148  &  63     & 13        & 229       \\
FAILED   &  134   &  10     & 8         & 76          \\
CANCELLED&   62   &  6      & 2         & 29        \\
TIMEOUT  &   17   &  41     & 9         & 235        \\
\end{tabular}
\caption{\textit{Power consumption per SLURM job status showing a significant portion of non completed Jobs.}}
\label{tab:slurm}
\end{table}
As a central point in machine learning, we investigate whether GPUs are used at their full capacity. As we can see in Fig.~\ref{fig:gpu}, most of the GPUs usages are not considering the full usage of the GPUs capacities from both the computing cores and the memory point of view.
This leaves the question whether these jobs could have been done faster, for instance by adapting the batch size or improving the pre-processing steps. Overall, these two findings confirm our hypothesis that users can benefit from recordings to optimize their efficiency and also validate the interest of such a simple profiling system.

\section{Conclusion}

This work presents a system to monitor energy behavior on a AI data center, with open source tools, further validated with external wattmeters. This setup can be quickly installed in most of the data centers. The behaviors we observed regarding the overall energy consumption show that GPU are under used and submitted jobs could be better configured. Given these results, we hypothesis that, even if the main environmental impacts come from other parts of the life cycle, indirect effects, or might concerns other criteria such as abiotic mineral depletion, approaching users with the topic of energy efficiency has an interesting potential as a first step to engagement, if properly contextualised.

\begin{acknowledgments}
  This work has been financed by "Réseau francilien en sciences informatiques" program.
\end{acknowledgments}

\bibliography{ms}

\end{document}